\documentclass[aps,prl,twocolumn,groupedaddress,amsmath,amssymb,showpacs]{revtex4}

\usepackage{graphicx}
\usepackage{hyperref}
\usepackage{pifont}

\begin{document}

\title{Resonant electronic states and $I$-$V$ curves of Fe/MgO/Fe(100) tunnel junctions}

\author{Ivan Rungger$^1$, Oleg N. Mryasov$^2$ and Stefano Sanvito$^1$}
\affiliation{$^1$ School of Physics and CRANN, Trinity College, Dublin 2, Ireland}
\affiliation{$^2$ Seagate Research, Pittsburgh, Pennsylvania, 15222, USA}

\date{August 6, 2008}

\begin{abstract}
The bias dependence of the tunnel magnetoresistance (TMR) of Fe/MgO/Fe tunnel junctions is investigated theoretically with a fully self-consistent scheme that combines the non-equilibrium Green's functions method with density functional theory. At voltages smaller than 20~mVolt the $I$-$V$ characteristics and the TMR are dominated by resonant transport through narrow interface states in the minority spin-band. In the parallel configuration this contribution is quenched by a voltage comparable to the energy width of the interface state, whereas it persists at all voltages in the anti-parallel configuration. At higher bias the transport is mainly determined by the relative positions of the $\Delta_1$ band-edges in the two Fe electrodes, which causes a decrease of the TMR.
\end{abstract}


\maketitle

Modern magnetic sensors, such as read heads for hard disk drives, are based on the tunnel magnetoresistance
(TMR) effect. This is the drop in resistance of a magnetic tunnel junction (MTJ) formed by two magnetic layers when
the mutual alignment of their magnetization vectors changes from antiparallel (AP) to parallel (P). The TMR magnitude
is given by $\mathrm{TMR}=[I_\mathrm{P}-I_\mathrm{AP}]/I_\mathrm{AP}$, with $I_\alpha(V)$ being the current
at the voltage $V$ for the $\alpha$ configuration (P or AP). Huge TMR ratios have been recently achieved in 
epitaxial, all crystalline Fe/MgO \cite{Yua2} and CoFeB/MgO \cite{Park1} MTJs, reaching up to 500\% at room 
temperature and 1010\% at 5~K \cite{Ohno1}.
These large values of TMR are broadly attributed to the phase coherent and transverse momentum conserving 
transport. The TMR thus is governed not only by the spin-polarization of the electrode density of states (DOS), 
but also by the details of the wave-functions matching across the barrier. Magnetic transition metals with bcc crystal 
structure (Fe) possess a high transmission $\Delta_1$ band, which decays slowly across the MgO barrier
\cite{But1}. This is fractionally filled for majority spins ($\uparrow$), and empty for minority spins
($\downarrow$). Since these bands dominate the tunneling current, bcc transition metals with MgO barrier
effectively behave as half-metals, and the TMR is expected to be very large \cite{But1,Math}.

Another important, but much less investigated aspect, is the relation between electronic states and 
the $I$-$V$ characteristics in these highly crystalline MTJs. Interfacial states and details of the Fe 
band-structure, otherwise washed out by disorder, play an important role in the transport and indeed can
be identified through the $I$-$V$ curves and its derivatives ($G(V)=\mathrm{d}I/\mathrm{d}V$ and
$S(V)=\mathrm{d}^2I/\mathrm{d}V^2$). For instance, high-quality conventional MTJs (2-3 nm MgO thickness) 
show a pronounced broad peak on the $S(V)$ curve at about 1~Volt for the AP, and a number of
small peaks at lower voltages in the P configuration \cite{Yua3,Tiusan:PRB08}. Combined with a quantitative theory these
measurements can provide a wealth of information, and help on the level of device design.
 We note that very thin MTJs, with a MgO thickness of about 1 nm, have to be used for ultra high recording density
($>$500~Gbit/inch$^2$) HDD readers \cite{Yua1}. Remarkably, at these small MgO thicknesses, the growth mode of the
Fe/MgO/Fe stack changes \cite{Tiusan:PRB08}.

In this letter we investigate theoretically the $I$-$V$ characteristics of perfectly crystalline Fe/MgO MTJs, and demonstrate that its features originate from the sweeping of the $\Delta_1$ band-edges and of interface states across the bias window.
These generic features are emphasized here, as they resemble closely those studied in molecular devices \cite{NatMat} 
and magnetic point contacts \cite{MPC}. We investigate the ultra-thin MTJ regime, with the goal to provide a solid basis 
for decoding future $I$-$V$ measurements of $\lesssim1$ nm thick MgO barriers, in term of their relation to the underlying 
electronic structure and possibly growth defects.
Calculations are performed using SMEAGOL \cite{Smeagol}, which combines the non-equilibrium
Green's function method with density functional theory (DFT) \cite{Siesta}. The total transmission coefficient
$T(E;V)$ is self-consistently evaluated at finite bias and integrated to give the spin-current.
\begin{equation}
I^\sigma(V)=\frac{e}{h}\int\mathrm{d}E\;T^\sigma(E;V)\left[f_+-f_-\right]\;.
\end{equation}
Here $\sigma$ is the spin index ($\uparrow$, $\downarrow$) and $f_\pm$ is the Fermi function calculated at
$(E-E_\mathrm{F}\pm eV/2)/k_\mathrm{B}\tau$, with $E_\mathrm{F}$ the Fe Fermi energy, $k_\mathrm{B}$ the
Boltzmann constant and $\tau$ the electronic temperature. Translational invariance allows to write
$
T^\sigma(E;V)=\frac{1}{\Omega_k}\int\:T^\sigma(E, \mathbf{k}_\perp;V)\;d{\mathbf{k}_\perp}\;,
$
where the integral runs over the 2D Brillouin zone perpendicular to the transport direction with area $\Omega_k$.
Large $\mathbf{k}_\perp$-point samplings are necessary to converge $T(E;V)$, so that an extremely stable
numerical algorithm is needed \cite{ivanselfenergy}.

We consider a MTJ oriented along the Fe(100) direction and formed by 4 MgO monolayers ($\sim$1.1~nm long). The
atomic coordinates are those from reference \cite{bluegel1}, which have been optimized by high-accuracy
plane-wave calculations. The unit cell used for the transport contains also 8 Fe atomic layers on each side
of the MgO barrier, which are enough to screen the charge density from the interface. A 7x7 $\mathbf{k}_\perp$-points mesh
converges the charge density, but a 100x100 mesh is used for evaluating $T(E;V)$. The basis set
employed is single-$\zeta$ for the Fe $p$ and $d$ orbitals, while double-$\zeta$ is used for all other orbitals \cite{Siesta}.
The local density approximation (LDA) is adopted throughout, and we use a real space mesh cutoff of 600~Ry
and $\tau=300$~K.

\begin{figure}
\center
\includegraphics[width=8.5cm,clip=true]{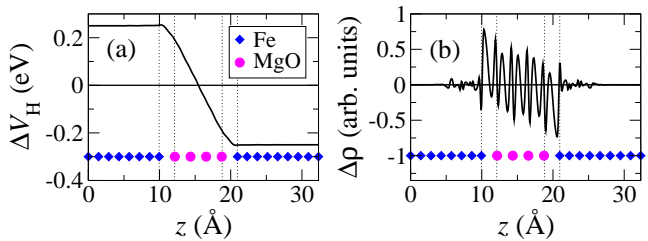}
\caption{(Color online) (a) Planar average $\Delta V_\mathrm{H}$ of the difference between the Hartree potential at 0.5
Volt and the one at 0-bias. (b) Planar average $\Delta \rho$ of the difference between the charge density at
0.5 Volt and the one at 0-bias. The diamonds and dots indicate the location of the Fe and MgO layers.}
\label{Fig0}
\end{figure}
First, we look at the electrostatic Hartree potential drop across the junction. In Fig. \ref{Fig0}(a) we show the difference $\Delta V_\mathrm{H}$ between the planar average of the self-consistent Hartree potential at a finite bias and that at 0-bias along the junction stack ($z$-axis). $\Delta V_\mathrm{H}$ is flat in the electrodes and decays linearly in the MgO, demonstrating that the usual approximation of shifting rigidly the electronic structure of the electrodes and then applying a linear potential drop across the barrier \cite{Mertig} is well justified.
In Fig. \ref{Fig0}(b) we also show the difference $\Delta \rho$ between the planar average of the self-consistent charge density at a finite bias and at 0-bias, so that $\Delta \rho(z)\propto - d^2\Delta V_\mathrm{H}(z)/dz^2$. At all voltages we find $\Delta \rho(z)$ increasing linearly with $V$, and charge accumulating at the extremal layers of the electrodes just before MgO. Inside the MgO $\Delta \rho(z)$ oscillates due to the electric field induced polarization. This is confirmed by a DFT calculation for an isolated MgO slab of the same thickness in an equivalent electric field, which shows analogous oscillations.

\begin{figure}
\center
\includegraphics[width=7.0cm,clip=true]{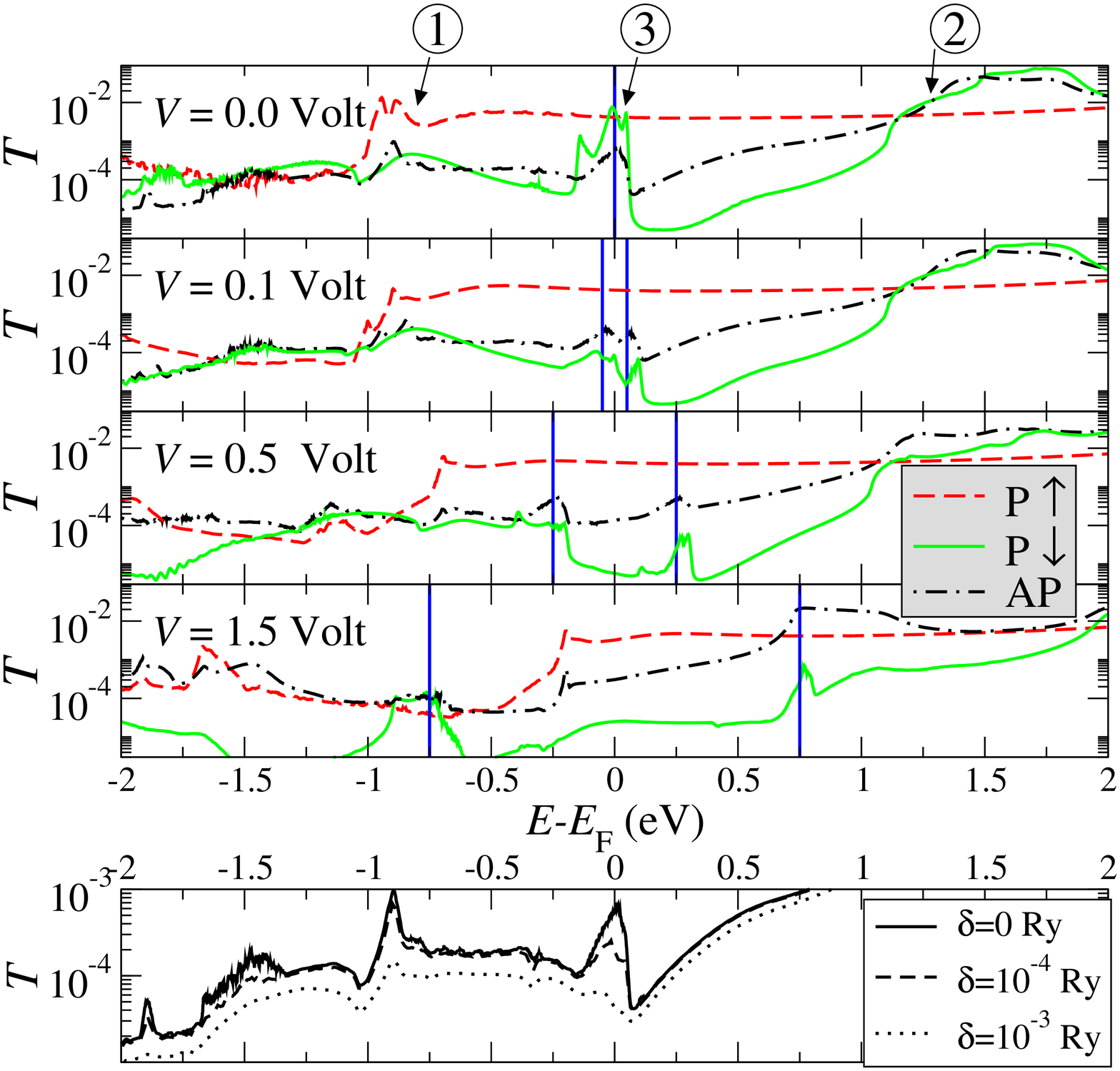}
\caption{(Color online) Transmission coefficient $T(E;V)$ as a function of energy, $E$, and for different biases, $V$.
The vertical lines are placed at $E=E_\mathrm{F}\pm eV/2$ and enclose the bias window.
The dashed (solid) line is for the $\uparrow$ ($\downarrow$) spin band in the P configuration, while the dashed-dotted
line is for the AP configuration. The lowest panel is $T(E;0)$ for the AP configuration and different broadening $\delta$.}
\label{Fig1}
\end{figure}

We start our analysis of the transport properties by presenting $T(E;V)$ (Fig.~\ref{Fig1}).
Three main features appear in the $V=0$ transmission coefficient: \ding{192} a sharp increase (note the logarithmic scale) in
transmission at around -1~eV for the $\uparrow$ spins in the P configuration, \ding{193} a similar, although smoother increase
above +1~eV for the $\downarrow$ spin in the P configuration and for the AP configuration, and \ding{194} a sharp resonance at
$E_\mathrm{F}$ for the $\downarrow$ spins in the P configuration, which is still visible in the AP configuration.

\begin{figure}
\center
\includegraphics[width=8.5cm,clip=true]{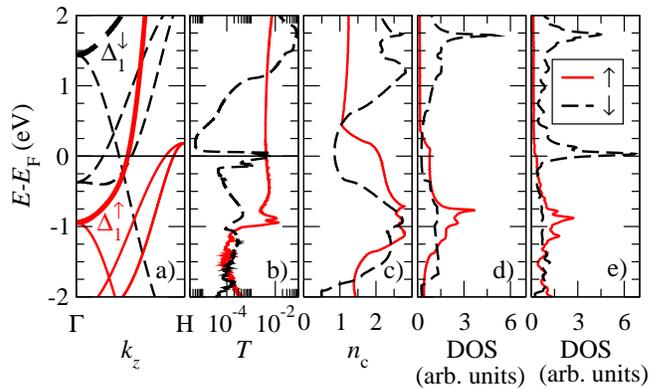}
\caption{(Color online) (a) Bulk Fe band-structure along the $\Gamma\rightarrow\mathrm{H}$ direction (the bands with $\Delta_1$-symmetry are emphasized), (b) $T(E;0)$ in the P configuration, (c) average number
of channels per $k$-point for bulk Fe, $n_\mathrm{c}$, (d) bulk Fe DOS, (e) interface Fe-layer DOS. Note that the $\Delta_1$ band-edges coincide with a sharp increase in the transmission coefficient $T(E;0)$ and with a large $n_\mathrm{c}$.}
\label{Fig2}
\end{figure}

The first two features are associated with the $\Delta_1$ band-edges, respectively for the $\uparrow$ and $\downarrow$ spins, as it can be clearly seen in Fig.~\ref{Fig2}. Note also that these band-edges coincide with energy regions where the average number of Fe-channels per $\mathbf{k}_\perp$-point $n_\mathrm{c}$ is maximized [Fig.~\ref{Fig2}(c)]. Since these
are a feature of the electronic structure of Fe alone, their position as a function of bias is set by the quasi Fermi energies, $E_\mathrm{F}\pm eV/2$, of the two electrodes. For instance the sharp increase of $T^\uparrow(E;V)$ in the P configuration (\ding{192}) moves to -0.75~eV and -0.25~eV respectively for voltages of 0.5~Volt and 1.5~Volt, following the $\Delta_1^\uparrow$ band-edge of the left lead. In the same way the broad peak in both P and AP at about 1.5 eV above $E_\mathrm{F}$ (\ding{193}) splits into two broad sub-peaks, separated by $eV$. Note however that whereas the height of the peak is roughly constant for the AP configuration, in the P configuration the sub-peak entering the bias window shrinks drastically.

\begin{figure}
\center
\includegraphics[width=8.0cm,clip=true]{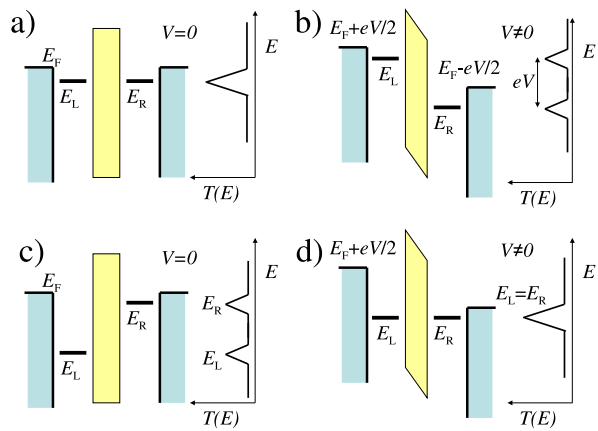}
\caption{(Color online) Schematic representation of the bias dependence of surface states localized at the interfaces between the magnetic electrodes and the tunnel barrier. (a) and (b) are for the left (L) and right (R) interface states having the same energy and therefore resonating at V=0; (c) and (d) are for the L and R interface states not resonating at zero-bias.}
\label{Fig3}
\end{figure}

In contrast the peak in transmission at $E_\mathrm{F}$ for the $\downarrow$ spins in the P configuration and in the AP configuration cannot be associated with leads band-edges but instead originates from a narrow interface state in the minority band \cite{IvanJEMS,But1,Tsym1}. This is spatially localized at the interface between Fe and MgO and it is resonating at $E_\mathrm{F}$. A comparison between the bulk Fe DOS and the DOS for the interface Fe layer [Figs.~\ref{Fig2}(d) and (e)] indicates the surface state for $\downarrow$ spins at $E_\mathrm{F}$ \cite{But1}.
The dynamics of such a interface state under bias can be
understood from the cartoon of figure \ref{Fig3}.
Consider the panels (a) and (b) where two identical interface states are localized on either sides of the tunnel barrier. This is the situation encountered here for the $\downarrow$ spins. The transport is then resonating across the barrier at the interface state energy $E_\mathrm{R}=E_\mathrm{L}$. In the case considered here the resonance energy is close to $E_\mathrm{F}$.

Since, in general the state is coupled more strongly to one of the electrodes, it will trace closely its quasi
Fermi energy. For instance for positive bias and a interface state localized on the left-hand (right-hand) side of
the junction [Fig.~\ref{Fig3}(b)] we obtain $E_\mathrm{L}(V)=E_\mathrm{L}(0)+eV/2$
($E_\mathrm{R}(V)=E_\mathrm{R}(0)-eV/2$). This brings
the states on either side of the junction out of resonance and generally suppresses the transmission. Thus the
peak in $T(E;0)$, originating from a resonating interface state across the barrier, will evolve into two smaller peaks
separated by an energy $eV$. Indeed this is the behavior observed in
Fig.~\ref{Fig1}. For instance the two peaks centered at $E_\mathrm{F}$ are separated by 0.5~eV and 1.5~eV
respectively for voltages of 0.5~Volt and 1.5~Volt. In the AP configuration this situation is found
at any bias, since the spin $\downarrow$ interface state is always present on one side of the junction only.

A second possible situation is when the interface states on the left-hand and right-hand side of
the tunnel junction have a different origin and are placed at different energy. In this case we do not expect
zero-bias resonance, however there will be a critical voltage at which the resonant condition is met. In this
case we expect the rise of a large peak in the transmission coefficient at a bias $eV=E_\mathrm{L}(0)-E_\mathrm{R}(0)$.
This situation has never been encountered for the symmetric MTJ investigated here, but it is likely to be the
most typical case in real junctions.

\begin{figure}
\center
\includegraphics[height=6.7cm,clip=true]{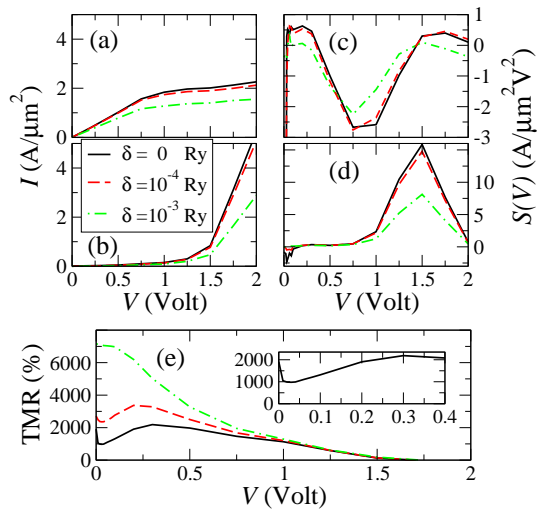}
\caption{(Color online) Current $I$, $S(V)$ and TMR as function of voltage $V$ for different values of the
imaginary part of the energy $\delta$: (a) $I$-$V$ for P, (b) $I$-$V$ for AP, (c) $S(V)$
for P, (d) $S(V)$ for AP, (e) TMR. In the inset of panel (e) we show the TMR in the low-bias
region for $\delta=0$.}
\label{Fig4}
\end{figure}

We now move to analyzing the $I$-$V$ characteristic and the TMR (Fig. \ref{Fig4}), starting from the
low bias region ($V<$ 0.4~Volt). The most apparent feature in this bias range is a sharp reduction of the TMR from its zero-bias value followed by a rapid increase which peaks at $V\approx0.3$~Volt. The sharp reduction can be associated with a decrease in the P $\downarrow$ current originating from the loss of the interface state resonant condition at $V\approx20$~mVolt, a bias which roughly corresponds to the line-width of the interface state. For $V<$~20~mVolt the P current is shared by the two spin-species, while for $V>$~20~mVolt the $\uparrow$ component dominates. Such a reduction in the P $\downarrow$ component at the bias corresponding to the resonant condition loss can be clearly observed in the $S(V)$ plot of Fig. \ref{Fig4}(c). As the bias further increases the conductance in the P configuration is approximately constant. In contrast the conductance of the AP configuration is reduced for $V$ between 0.1 and 0.3~Volt, and this behavior results in an increase of the TMR with the broad peak at 0.3 Volt.

The high bias region is characterized by a decrease of the conductance in the P configuration at around 1~Volt
and a dramatic increase of the AP current for $V>1.5$~Volt. This produces a strong reduction of the TMR with bias
and an almost complete suppression for $V>1.5$~Volt. In fact at about 1.75~Volt the TMR becomes negative. Such
a voltage range should be put in comparison with the band offset. In our LDA calculations $E_\mathrm{F}$ for Fe is
positioned $\sim$1.8~eV below the MgO conduction band minimum and $\sim$3.0~eV above the valence band
maximum. This means that voltages of the order of 1.5~Volt are still rather far from those needed
for tunneling across a reduced barrier.

The high-bias behavior is dominated by the relative energy shift with bias of the Fe $\Delta_1^\sigma$ states.
The origin of the reduction of the conductance in the P configuration for $V>$~1~Volt is that once the $\Delta_1^\uparrow$ band-edge of the
left lead enters the bias window, the high transmission region only extends over part of such window
(Fig. \ref{Fig1}), so that the increase of the current with bias is reduced by a factor of about 2. Note that the current in the $\downarrow$ is negligible.
The broad peak of the $\downarrow$ transmission at about 1.5 eV above the Fermi energy, related to the $\Delta_1^\downarrow$ band-edge, never contributes significantly to the current in the P configuration. The reason is that whereas for positive bias the $\Delta_1^\downarrow$ band-edge of the right lead enters the bias window, the one on the left lead moves away from the bias window, so that the peak shrinks as bias is applied. The general evolution with bias of this transmission peak resembles the one of the surface states schematically shown in Figs. \ref{Fig3}(a) and (b).
In the AP configuration however this peak does contribute to the current, since electrons belonging to the $\Delta_1^\downarrow$ band in the
right lead that are inside the bias window can tunnel into the $\Delta_1^\uparrow$ bands of the left lead. This
leads to an increase of the conductance once the $\Delta_1^\downarrow$ band-edge moves into the bias window, resulting
in a drastic increase of the AP current at $V>1.5$~Volt.
We note that our results in the high bias region are in contrast with those in reference \cite{guo2}, in which a rapid increase
of the conductance is found also for the P configuration.

It is also interesting to comment on the $S(V)$ plots [Fig.\ref{Fig4}(c) and (d)]. For both the magnetic configurations
one can observe a peaked structure. This is observed also experimentally \cite{Yua3} and attributed to resonances
in the transmission. At the qualitative level this interpretation is confirmed by our results in the low bias region, where
the interface states determine the behavior of the $I$-$V$. The actual position of the peaks however depends on the
atomic details of the two interfaces and we do not expect agreement. The high-bias region however is more
problematic. In particular we note that our $I$-$V$ for the P configuration does not increase as rapidly as that found
in typical experiments. One possible explanation for such difference is that the MgO barrier calculated here is very
thin and the current consequently already large at small bias, another reason could be presence
of defects at the interface and in the MgO in experiments.

Finally we investigate the effect of generic disorder, motivated by observations \cite{Tiusan:PRB08}. This is modeled at a simple level by adding a small imaginary part $\delta$ to the energy when calculating $T(E;V)$, i.e. it corresponds to a uniform level broadening \cite{Tsym1}. The effects of such addition on the $I$-$V$, $S(V)$ and TMR are shown in Fig.~\ref{Fig4}, while those on $T(E;0)$ for the AP configuration are shown in Fig.~\ref{Fig1}. Fig. \ref{Fig1} clearly shows that increasing $\delta$ results in a gradual suppression of the $T(E;0)$ resonance at $E_\mathrm{F}$. For $\delta=10^{-4}$~Ry the peak is still clearly visible, and only for $\delta=10^{-3}$~Ry it completely disappears. As a result the TMR at low bias largely increases and the non-monotonic behavior for $V<0.4$~Volt is suppressed. In contrast the high bias region is barely affected by $\delta$. These results, although indicative of the effects of disorder on the TMR, should be taken with caution. The broadening $\delta$ introduces unstructured disorder, and the transmission of all the spin-channels is equally reduced. In reality one may expect the transmission to either increase or decrease depending on the type of scattering center, which in general will act differently on the different spin-channels. This will in general result in an enhancement of the current in the AP alignment, causing a reduction of the TMR.

In conclusion, we investigated the bias dependence of the TMR of an epitaxial Fe/MgO/Fe tunnel junction. We identify two different bias regions, which are affected by two types of electronic resonances: (i) interface resonance states and (ii) band edges. The low bias region is characterized by a non-monotonic behavior of the TMR caused by resonant scattering across the barrier due to the highly localized interface states. In the high bias region ($V>0.4$~Volt) the TMR decreases monotonically, mainly due to band edge states. We have also shown how disorder can suppress the transport through interface states and smoothens the TMR.

This work is sponsored by Science Foundation of Ireland (
07/IN.1/I945 and 07/RFP/PHYF235) and by Seagate. Computational
resources have been provided by TCHPC and ICHEC. O.M. acknowledges the 
SFI C.T.Walton fellowship support during his stay at Trinity College.

\end{document}